%% file: main.tex
\documentclass{PoS}

\usepackage{amsmath}
\usepackage{amssymb}
\usepackage{braket}
\usepackage{cite} 
\usepackage[utf8]{inputenc}
\usepackage{graphicx}
\usepackage{tej}
\usepackage{todonotes}
\usepackage{esint}

\input{defines.tex}

\author{William Detmold, \speaker{Gurtej Kanwar\speaker{} and Michael L. Wagman}\\
  Center for Theoretical Physics,
  Massachusetts Institute of Technology,
  Cambridge, MA 02139, USA\\
  E-mail: \email{wdetmold@mit.edu},
  \email{gurtej@mit.edu}, \email{mlwagman@mit.edu}
}

\abstract{
  Correlation functions in one-dimensional complex scalar field theory provide a toy model for phase fluctuations, sign problems, and signal-to-noise problems in lattice field theory.
  Phase unwrapping techniques from signal processing are applied to lattice field theory in order to map compact random phases to noncompact random variables that can be numerically sampled without sign or signal-to-noise problems.
  A cumulant expansion can be used to reconstruct average correlation functions from moments of unwrapped phases, but points where the field magnitude fluctuates close to zero lead to ambiguities in the definition of the unwrapped phase and significant noise at higher orders in the cumulant expansion.
  Phase unwrapping algorithms that average fluctuations over physical length scales improve, but do not completely resolve, these issues in one dimension.
  Similar issues are seen in other applications of phase unwrapping, where they are found to be more tractable in higher dimensions.
}

\title{Unwrapping phase fluctuations in one dimension}

\ShortTitle{Unwrapping phase fluctuations in 1D}

\FullConference{The 36th Annual International Symposium on Lattice Field Theory - LATTICE2018\\
		22-28 July, 2018\\
		Michigan State University, East Lansing, Michigan, USA.}

\begin{document}

\section{Introduction}

Lattice quantum field theory (LQFT) can be used to predict the properties of strongly interacting systems directly from the Standard Model or from low-energy effective field theories.
Path integrals are approximated as finite-dimensional integrals in LQFT that can sometimes be efficiently computed with Monte Carlo (MC) methods.
However, for many interesting systems ranging from nuclei and neutron stars to strongly correlated electrons, LQFT path integrals face sign problems that obstruct efficient MC methods using importance sampling.

Sign problems arise when different field configurations make contributions to path integrals that have different signs or phases.
When a sign problem is present, the integrand of the path integral cannot be interpreted as a probability distribution and importance sampling cannot be used.
Instead, one can MC sample according to a different probability distribution and then reweight the contribution of each field configuration by the ratio of the integrand to the distribution used for sampling.
In reweighting approaches to the baryon chemical potential sign problem that use the zero-density lattice quantum chromodynamics (LQCD) partition function for importance sampling, the signal-to-noise (StN) ratio of this reweighting factor vanishes exponentially as the spacetime volume is taken to infinity~\cite{Gibbs:1986ut,Cohen:2003kd,Splittorff:2006fu,deForcrand:2010ys,Alexandru:2014hga}.
Standard calculations of hadronic correlation functions similarly involve averages of zero-density importance-sampled gauge field configurations that are weighted by the value of the correlation function in each configuration.
The StN ratio for baryon correlation functions decreases exponentially as the baryon number or source/sink separation are increased with a rate predicted by the moment analysis of Parisi~\cite{Parisi:1983ae} and Lepage~\cite{Lepage:1989hd}.
The baryon StN problem arises from phase fluctuations between correlation functions in different gauge field configurations and is therefore another manifestation of the baryon number sign problem~\cite{Wagman:2016bam}.

This work considers zero-plus-one-dimensional [$(0+1)D$] complex scalar field theory as a toy model for StN problems arising from phase fluctuations where possible solutions can be tested on an analytically tractable system.
The distribution of phase fluctuations in complex scalar field correlation functions with nonzero $U(1)$ charge is found to qualitatively resemble the distribution of LQCD baryon correlation function phase fluctuations described in Ref.~\cite{Wagman:2016bam}.
In an analytically tractable approximation, complex scalar field phase fluctuations are shown to be wrapped normally distributed with an exponentially-severe StN problem.
Building on the idea that StN problems arise whenever phase fluctuations are sampled numerically, this work explores a new method in which
 phase differences are ``unwrapped,'' or numerically integrated over a series of spacetime separations.
The resulting unwrapped phases are noncompact random variables rather than circular random variables defined modulo $2\pi$.
Moments of unwrapped phase differences can be calculated from positive-definite path integrals that do not have sign problems and do not generically require computational resources that increase exponentially with increasing $U(1)$ charge.
Correlation functions can be calculated from moments of unwrapped phase differences using cumulant expansion techniques similar to those of Ref.~\cite{Endres:2011jm}.\footnote{Cumulant expansions of noncompact ``extensive phases'' have also been applied to sign problems in QCD and other theories at nonzero chemical potential~\cite{Ejiri:2007ga,Nakagawa:2011eu,Ejiri:2012wp,Greensite:2013gya,Garron:2017fta,Bloch:2018yhu}.}
The phase unwrapping techniques used here are analogous to phase unwrapping techniques used elsewhere in science and engineering including signal processing, radar interferometry, x-ray crystallography, and magnetic resonance imaging~\cite{Judge:1994,Ghiglia:98,Ying:2006,Kitahara:2015}.

Though the $1D$ phase unwrapping algorithms studied here are not specific to this toy model,
$1D$ phase unwrapping algorithms generically suffer from numerical instabilities and do not immediately provide a robust solution to LQFT sign and StN problems.
Multidimensional phase unwrapping algorithms are known to avoid analogous numerical instabilities in other contexts, and more robust phase unwrapping algorithms might be achieved in future investigations of phase unwrapping in multidimensional LQFTs.

\section{Complex scalar field statistics}\label{sec:scalarstats}

Consider a complex scalar field $\varphi(t)$ with a $U(1)$-invariant interaction $V(|\varphi(t)|)$ where $t=0,\dots,L-1$ is a uniform periodic lattice.
With lattice spacing set to unity, the Euclidean action is
 \begin{equation}
   \begin{split}
     S(\varphi)     &\equiv \sum_{t=0}^{L-1} \big( \varphi^*(t+1) - \varphi^*(t) \big)\big( \varphi(t+1) - \varphi(t) \big) + M^2 |\varphi(t)|^2 + V(|\varphi(t)|).
   \end{split}\label{eq:Sdef}
 \end{equation}
The action has a $U(1)$ symmetry, $\varphi \rightarrow e^{-i\alpha}\varphi$,
that can be used to classify sectors of states in the LQFT Hilbert space that do not mix under (Euclidean) time evolution.
Field products of the form $\mathcal{O}_{Q,2P}(t) \equiv \varphi(t)^{Q}|\varphi(t)|^{2P}$ for $Q\geq 0$ and $\mathcal{O}_{-Q,2P} \equiv \mathcal{O}_{Q,2P}^*$
transform under $U(1)$ in the charge $Q$ representation.
Green's functions of these operators,
\begin{equation}
   \begin{split}
      G_{Q,2P}(t) = \langle \mathcal{O}_{Q,2P}(t) \mathcal{O}^*_{Q,2P}(0) \rangle \equiv \sum_{n=1}^\infty Z_{n;Q,2P} e^{-E_{n;Q} t} \left[ 1 + O\paren{e^{-E_{n;Q}(L-t)}} \right],
   \end{split}\label{eq:GQ2Pdef}
\end{equation}
access all states of the theory; see Ref.~\cite{Detmold:PhaseUnwrapping} for more details.
The scalar boson propagator is $G(t) \equiv G_{1,0}(t)$ and in the non-interacting case $V=0$ its mass is $E \equiv 2\  \text{arcsinh}(M/2)$.

Since the action
is real, $e^{-S}$ is a positive-definite function that can be interpreted as a probability distribution
$    \mathcal{P}(\varphi) \equiv \frac{1}{Z} e^{-S(\varphi)}$.
MC methods can be used to produce stochastic samples from this distribution of complex scalar fields $\varphi_i$ and composite operators $\mathcal{O}_{Q,2P}^i$ with $i=1,\dots, N$.
Correlation functions can be approximated with ensemble averages,
\begin{equation}
   \begin{split}
      \overline{G}_{Q,2P} = \frac{1}{N}\sum_{i=1}^N \mathcal{O}_{Q,2P}^i(t) \sq{\mathcal{O}_{Q,2P}^i(0)}^* = G_{Q,2P}\left[ 1 + O\left( \frac{\text{Var}(G_{Q,2P}) }{\sqrt{N}}\right) \right].
   \end{split}\label{eq:Gbardef}
\end{equation}
Ground-state energies are approximated by 
$E_{Q,2P}(t) \equiv - \ln\left( \overline{G}_{Q,2P}(t+1) \right) + \ln\left( \overline{G}_{Q,2P}(t) \right)$ as $t\rightarrow \infty$.

The StN problem can be analytically estimated for the non-interacting theory. Following standard Parisi-Lepage arguments~\cite{Parisi:1983ae,Lepage:1989hd}, the variance of $G_{Q,2P}$ can be described by a linear combination of correlation functions.
The variance of $\overline{G}_{Q,2P}$ is related to the variance of $G_{Q,2P}$ by $1/\sqrt{N}$ in the large-$N$ limit and ignoring finite-$L$ effects is given by
\begin{equation}
  \begin{split}
    \text{StN}(\text{Re}[\overline{G}_{Q,2P}(t)]) &\equiv \frac{G_{Q,2P}(t)}{\sqrt{\text{Var}(\text{Re}[\overline{G}_{Q,2P}(t)])}} 
    = \sqrt{2 N} e^{-Q E t} \left[ 1 + O\left( e^{-2 E t} \right) + O(N^{-1/2}) \right].
  \end{split}\label{eq:GPQStN}
\end{equation}
Correlation functions describing sectors with $U(1)$ charge $Q\neq 0$ face an exponentially severe StN problem where the exponent is proportional to the $U(1)$ charge of the system.

\begin{figure}[t]
\centering
\includegraphics[width=\textwidth]{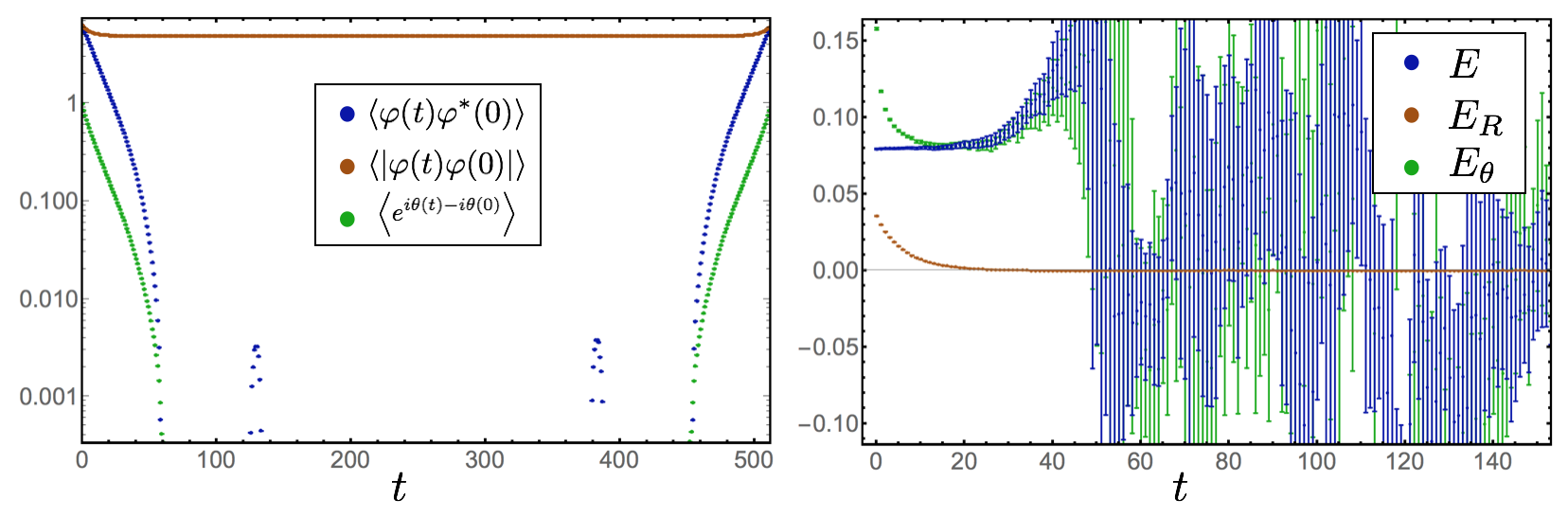}
\caption{The left plot shows a magnitude-phase decomposition of the scalar boson propagator for the non-interacting MC ensemble $C_0$ described in Sec.~\ref{sec:unwrap1D} with ensemble averages of the propagator, propagator magnitude, and propagator phase in blue, orange, and green respectively. The right plot shows the corresponding effective masses $E = -\partial_t \ln \avg{\varphi(t)\varphi^*(0)}$, $E_R = -\partial_t \ln \avg{|\varphi(t)\varphi(0)|}$, and $E_\theta = -\partial_t \ln \avg{e^{i\Theta(t)}}$. This and other figures are reproduced from Ref.~\cite{Detmold:PhaseUnwrapping}.}
\label{fig:freeStN}
\end{figure}

Analogously to the phase fluctuations associated with baryon number charge~\cite{Wagman:2016bam},
phase fluctuations associated with $U(1)$ charge are responsible for the StN problem in the free scalar theory.
Fig.~\ref{fig:freeStN} shows a decomposition of correlator and effective mass into
magnitude and phase $\avg{e^{i\theta(t) - i\theta(0)}} \equiv \avg{e^{i\Theta(t)}}$ components. The scalar boson propagator magnitude is $O(1)$
both sample-by-sample and in expectation with no severe StN problem.
The phase of the scalar boson propagator is $O(1)$ sample-by-sample by definition but $O(e^{-Et})$ in expectation with a severe $O(e^{-Et})$ StN problem.
Analogous behavior occurs for generic correlation functions.
The phase of a general correlation function
$    \Theta_Q(t) \equiv \text{arg}\left[ \mathcal{O}_{Q,2P}(t)\mathcal{O}_{Q,2P}^*(0) \right] = iQ \Theta(t)$
depends on the $U(1)$ charge of the correlation function
and $\avg{e^{i\Theta_Q}}$ has both an expectation value and a StN problem of $O(e^{-Q E t})$.

For arbitrary $V(|\varphi|)$, the partition function factors into magnitude and phase contributions,
\begin{equation}
  \begin{split}
    Z &= \int_0^\infty \prod_{t= 0}^{L-1} \left[
      d|\varphi(t)|\; |\varphi(t)| \; e^{-2|\varphi(t)|^2 - V(|\varphi(t)|)} \right]
    \int_{-\pi}^\pi \prod_{t= 0}^{L-1} \left[
      \frac{1}{\pi }d \theta(t) \; e^{\kappa(t) \cos(\Delta(t))} \right],
  \end{split}\label{eq:Zdecomp}
\end{equation}
with
$ \kappa(t) \equiv 2 |\varphi(t)||\varphi(t-1)| $ and 
$ \Delta(t) \equiv \theta(t) - \theta(t-1)$.
For a given scalar field magnitude the phase differences $\Delta(t)$
are independent in the $L\rightarrow \infty$ limit where the PBC constraint $\sum_{t=0}^{L-1} \Delta(t) = 2\pi w$ can be neglected.
The $L\rightarrow \infty$ distribution for $\Delta(t)$ is a von Mises
distribution and is well studied in circular statistics~\cite{Fisher:1995,Mardia:2009},
    $ \mathcal{P}(\Delta(t)) =  
e^{\kappa(t)\cos(\Delta(t))} / 2\pi I_0(\kappa(t))$.

\begin{figure}[t]
\centering
\includegraphics[width=\textwidth]{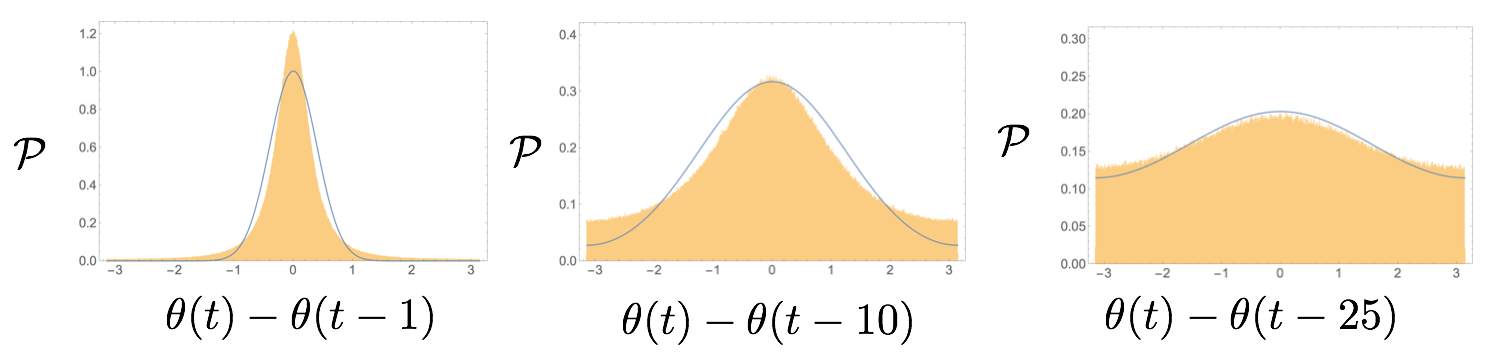}
\caption{ Histograms of differences of correlator phase differences at separations 1, 10, and 25 lattice sites for ensemble $C_0$.
  The histograms represent the empirical probability distribution functions $\mathcal{P}(\theta(t) - \theta(t-\delta t))$ with $\delta t = 1,\ 10,$ and $25$ respectively, while the blue curves show wrapped normal fits predicted by Eq.~\eqref{eq:WNtheta}.
  Heavy tails not reproduced by the wrapped normal fits are apparent and also arise for baryon phases in LQCD~\cite{Wagman:2016bam} as well as for the real parts of baryon correlation functions~\cite{davidkaplanLuschertalk}.
}
\label{fig:phasehist}
\end{figure}

It is difficult to calculate further properties of this distribution analytically.
Instead, the analysis is further simplified by assuming small magnitude fluctuations and phase differences,
\begin{equation}
  \begin{split}
    \frac{|\varphi(t)||\varphi(t^\prime)| - \avg{|\varphi(t)||\varphi(t^\prime)|}}{\avg{|\varphi(t)||\varphi(t^\prime)|}} \ll 1    
	\qquad \text{and} \qquad
	\Delta(t) \ll 1.
  \end{split}\label{eq:small}
\end{equation}
For fine discretizations with $M^2 \ll 1$, the gradient term provides the dominant contribution to the action and Eq.~\eqref{eq:small} should approximately hold for generic neighborhoods of generic field configurations.
In this approximation, phase differences between adjacent lattice sites are identically distributed as well as independent since
    $ \kappa(t) \approx \kappa \equiv \frac{1}{L} \sum_{t} \avg{\kappa(t)}$.
The distribution of $\Delta$ becomes
\begin{equation}
  \begin{split}
    \mathcal{P}(\Delta) &\approx \sqrt{\frac{\kappa}{2\pi}}\sum_{k\in\mathbb{Z}}  e^{-\kappa (\Delta + 2\pi k)^2 / 2} 
    = \frac{1}{2\pi}\sum_{n\in\mathbb{Z}} e^{i n \Delta-n^2 /(2\kappa)},
  \end{split}\label{eq:WN}
\end{equation}
a wrapped normal distribution 
describing a normally-distributed random variable defined mod $2\pi$.
This approximation is compared to empirical distributions of the correlator phase in Fig.~\ref{fig:phasehist}.

The wrapped normal characteristic function is identical to the normal characteristic function,
$ \Phi_{\mathcal{P}(\Delta)}(n) \equiv \avg{e^{in\Delta}} \approx   e^{-n^2 /2\kappa}$.
The characteristic function for propagator phase $\Theta$ can be described as a product of characteristic functions of $\Delta$,
such that Fourier transformation gives
\begin{equation}
  \begin{split}
    \mathcal{P}(\Theta) &= \frac{1}{2\pi}\sum_{n\in\mathbb{Z}} e^{-in\Theta} \Phi_{\mathcal{P}(\Theta)}(n) 
    \approx \frac{1}{2\pi}\sum_{n\in\mathbb{Z}} e^{-in \Theta} e^{ - t n^2 /2\kappa}.
  \end{split}\label{eq:WNtheta}
\end{equation}
Under the assumptions of Eq.~\eqref{eq:small}, the scalar boson propagator is given by
$G(t) \approx \avg{|\varphi(t) \varphi(0)|} \times  e^{-t/(2\kappa)}$
with $\kappa \approx 1 / 2E$ and
$\avg{|\varphi(t)\varphi(0)|} \approx  Z_{1;0,1}$.
The expectation value of the ensemble-average correlation function can be calculated in this approximation as $\avg{\overline{G}} \approx Z_{1;0,1} e^{-E t}$.
Its variance is given by
$    \text{Var}(\overline{G})  
\approx Z_{1;0,1}^2\left(1 - e^{-2Et} \right) / 2N$,
and its StN ratio is
\begin{equation}
  \begin{split}
    \text{StN}(\overline{G}) &= \frac{\avg{\overline{G}}}{\sqrt{\text{Var}(\overline{G})}} \approx \sqrt{2N}  e^{-Et} \sq{1 + O\paren{e^{-Et}}}.
  \end{split}\label{eq:GbarStN}
\end{equation}
The full StN problem for the scalar propagator arises even under the approximations of Eq.~\eqref{eq:small}.
Determination of the  scalar propagator pole mass by MC sampling phases distributed according to Eq.~\eqref{eq:WNtheta} is equivalent to parameter inference for a wrapped normal distribution with variance $1/\kappa \approx 2E$.
Avoiding large finite sample size errors in wrapped normal parameter inference requires
    $ \frac{1}{\sqrt{N}} \lesssim \avg{\cos(\Theta)} \approx e^{-E t} $~\cite{Fisher:1995}
indicating that the window of time in which reliable parameter inference is possible has size scaling only as $\log{N}$.

\section{Unwrapped phase statistics}\label{sec:unwrappedstats}

The analysis above suggests that avoiding sign and StN problems is equivalent to avoiding numerical sampling of circular random variables. 
In this simple theory, the phase can be analytically integrated out to produce a dual theory with positive-definite path integral representations for correlation functions~\cite{Detmold:PhaseUnwrapping,Endres:2006xu}.
As a tractable alternative for more complicated theories where all phase variables cannot be integrated out analytically, one can imagine numerically sampling the real-valued angular displacement accumulated by the phase along $[0,t]$ including any $2\pi$ revolutions about the unit circle.
A variety of ``phase unwrapping'' techniques have been developed in other contexts to extract noncompact variables representing angular displacement from numerical samples of compact phases; see Refs.~\cite{Judge:1994,Ghiglia:98,Ying:2006,Kitahara:2015} for reviews.

An unwrapped phase defined by integrating phase differences along a $1D$ path satisfies 
\begin{equation}
  \begin{split}
    \widetilde{\Theta}(t) \equiv \widetilde{\theta}(t) - \widetilde{\theta}(0) \equiv \Theta(t) + 2\pi \nu(t).
  \end{split}\label{eq:ArgGnu}
\end{equation}
The unwrapped phase difference $\widetilde{\Theta}(t)$ associated with a LQFT propagator
is the principal-valued or ``wrapped'' phase difference plus
$2\pi$ times a winding number $\nu(t)$ equal to the total number of oriented branch cut crossings along the integration path.
The phase unwrapping problem is to determine winding numbers $\nu(t)$ that make the unwrapped phase $\widetilde{\theta}(t)$ a nearly continuous function of $t$ even across the branch cuts of $\theta(t)$.

\begin{figure}[t]
\centering
\includegraphics[width=\textwidth]{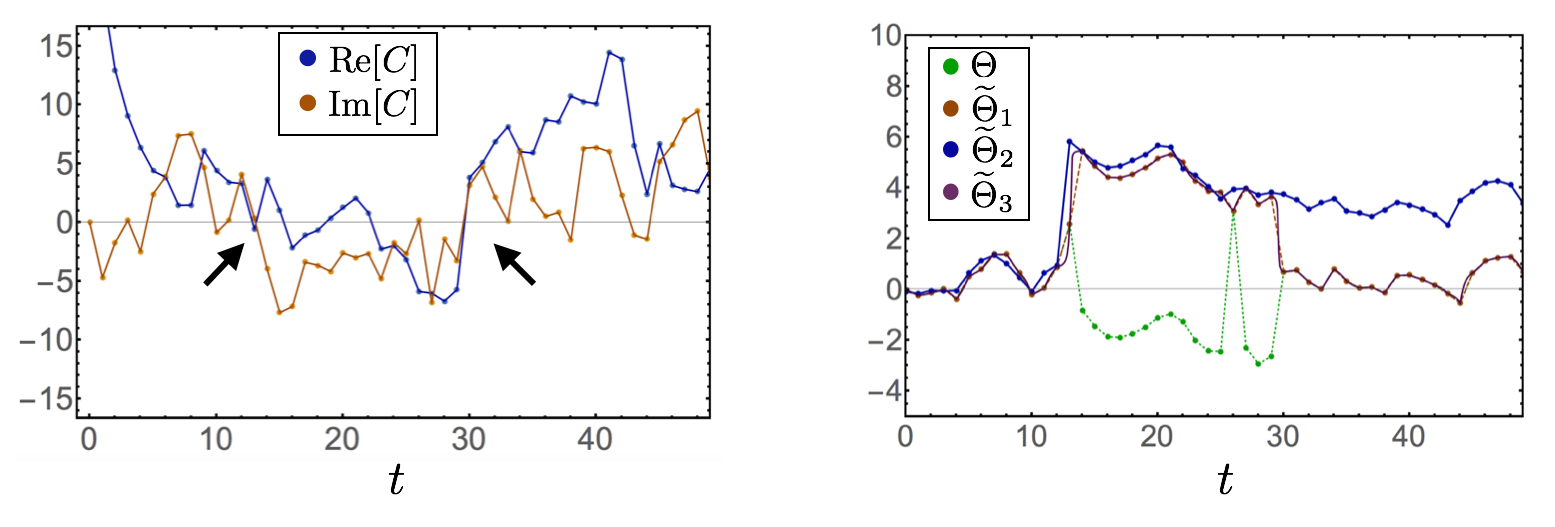}
\caption{The left plot shows real and imaginary parts of one MC sample of the $(0+1)D$ free complex scalar field propagator $C(t)$ exhibiting near-zeros of the magnitude indicated by the arrows. The right plot shows the wrapped correlator phase $\Theta(t) = \text{arg}C(t)$ and three definitions of the unwrapped phase denoted $\UWTheta_1$, $\UWTheta_2$, and $\UWTheta_3$, respectively produced by integrating a linear discretization of $d\widetilde{\theta}/dt$, assuming Eq.~\eqref{eq:1Dbound}, and algebraic phase unwrapping~\cite{Kitahara:2015} of a linear polynomial interpolation of $C(t)$.
  The numerical and algebraic winding number methods agree exactly at all lattice sites and only differ in their interpolation between lattice sites. Numerical integration 
  leads to $O(\pi)$ deviations of $\UWTheta_1$ from $\UWTheta_2$ and $\UWTheta_3$ for all $t > 30$.}
\label{fig:hardunwrapping}
\end{figure}

If one assumes a true unwrapped distribution arising from sampling a smooth function with sufficiently fine resolution, such that the unwrapped phases satisfy
\begin{equation}
  \begin{split}
    |\widetilde{\theta}(t) - \widetilde{\theta}(t-1)| < \pi,
  \end{split}\label{eq:1Dbound}
\end{equation}
Itoh demonstrated in Ref.~\cite{Itoh:1982} that a unique assignment of winding numbers $\nu(t)$
results.
In LQFT, quantum fluctuations can lead to points where  $|\varphi(t)| \approx 0$  for which $d\widetilde{\theta}/dt$ is nearly singular and Eq.~\eqref{eq:1Dbound} is not a good assumption, as shown in Fig.~\ref{fig:hardunwrapping}.
The violation of Eq.~\eqref{eq:ArgGnu} leads to $O(\pi)$ discrepancies between different winding number definitions at all lattice sites with $t$ larger than the site where Eq.~\eqref{eq:1Dbound} is violated.
Near-zeros therefore produce an accumulating $O(\pi)$ sensitivity in the unwrapped phase
which increases with increasing $t$.
This accumulation-of-errors problem is revisited in Sec.~\ref{sec:unwrap1D}.

All phase unwrapping schemes applied below meet the constraint that the unwrapped phase $\widetilde{\theta}$ differs from $\theta$ by $2\pi$ times an integer winding number $\nu$ and therefore that 
\begin{equation}
  \begin{split}
    W[\widetilde{\theta}] \equiv \widetilde{\theta} \text{ mod } 2\pi = \theta ,
  \end{split}\label{eq:Wdef}
\end{equation}
where the wrapping operator $W$ restricts the unwrapped phase to the interval $(-\pi,\pi]$.
This ensures that the average wrapped and unwrapped correlation functions are identical
\begin{equation}
  \begin{split}
    \avg{e^{\R +i\UWTheta}} &= \avg{e^{\R + i(\Theta + 2\pi\nu)}} 
    =  \avg{e^{\R + i\Theta}} 
    \approx Z_{1;0,1} e^{-E t}.
  \end{split}\label{eq:cosUW}
\end{equation}
Since wrapped phase differences $\Theta$ are related to $\UWTheta$ by applying $W$,
a normal distribution for $\UWTheta$ is compatible with the approximate wrapped normal distribution
for $\Theta$, Eq.~\eqref{eq:WN}.
Under these assumptions, the boson mass is given as $N\rightarrow \infty$ by
\begin{equation}
  \begin{split}
    \widetilde{E}(t) \equiv   \frac{1}{2N}\sum_{i=1}^N  \left[ \widetilde{\Theta}_i(t)^2 - \widetilde{\Theta}_i(t+1)^2 \right].
  \end{split}\label{eq:Mbartildedef}
\end{equation}
The StN ratio for the associated correlation function $\widetilde{G}(t)$ is
\begin{equation}
  \begin{split}
    \text{StN}\left( \widetilde{G}(t) \right) \approx \frac{\sqrt{N}}{\sqrt{2} Et}\left[ 1 + O(N^{-1/2})\right].
  \end{split}\label{eq:StNMbartilde}
\end{equation}
Eq.~\eqref{eq:StNMbartilde} demonstrates that normally-distributed unwrapped phases provide correlation function estimates whose StN ratios decrease polynomially as $t^{-1}$ rather than exponentially as $e^{-E t}$ as the spacetime volume $t$ containing nonzero $U(1)$ charge is increased.

For field configurations violating the small fluctuation assumptions of Eq.~\eqref{eq:small},
it is necessary to construct estimators for the correlator
$\langle e^{\R + i \UWTheta} \rangle$, and more generally Green's functions
$G_{Q,2P} = \langle e^{\R_{Q,2P} + i \UWTheta_Q} \rangle$,
that do not depend on assumptions about the distribution of $\R$ and $\Theta$.
Due to the integer winding number constraint, $W[\widetilde{\theta}] = \theta$,
the characteristic functions for the wrapped or unwrapped samples agree at $n = 1$,
    $ G_{Q,2P} = \Phi_{\R_{Q,2P} +i\Theta_Q}(1) = \Phi_{\R_{Q,2P} + i\UWTheta_Q}(1) $.
Once the unwrapped characteristic function is fit to numerical results by some method,
the Green's function can be estimated by evaluating the resultant fit function at $n=1$,
\begin{equation}
  \begin{split}
     \Phi_{\R_{Q,2P} + i\UWTheta_Q}(1) = \avg{e^{\R_{Q,2P} + i\UWTheta}} \approx \sum_n Z_{n;Q,2P} \, e^{-E_n t}.
  \end{split}\label{eq:avecos}
\end{equation}

Cumulant expansion methods similar to those explored in Refs.~\cite{Endres:2011jm,Endres:2011er,Grabowska:2012ik}  can be used to estimate $\Phi_{\R_{Q,2P} + i \UWTheta_Q}(1)$. 
The cumulants for a generic characteristic function $\Phi_z(k)$ are defined by the Taylor series for
$\ln(\Phi_z)$, and 
can be related to the moments of $z$. 
An estimator for the ensemble-average correlation function and ground-state energy can then be defined in terms of
cumulant estimates,
\begin{equation}
  \begin{split}
    \widetilde{G}_{Q,2P}^{(n_{max})} = \exp\left[ \sum_{n=1}^{n_{max}} \frac{1}{n!}\kappa_n \left( \R_{Q,2P} + i \widetilde{\Theta}_Q \right) \right] \hspace{20pt} \text{and} 
    \hspace{20pt} \widetilde{E}_{Q,2P}^{(n_{max})} = -\partial_t \log{\widetilde{G}_{Q,2P}^{(n_{max})}}.
  \end{split}\label{eq:EQPdef}
\end{equation}
 In the limits $n_{max}\rightarrow \infty$ and $N\rightarrow \infty$, the estimate $\widetilde{G}^{(n_{max})}_{Q,2P}$ should approach the average correlation function $G_{Q,2P}$ and for sufficiently large source/sink separation, the ground-state energy $\widetilde{E}^{(n_{max})}_{Q,2P}$ should approach $E_{Q,0}$.

The leading contributions to Eq.~\eqref{eq:EQPdef} are
    $\kappa_1(\R_{Q,2P})$, 
    $\kappa_2(\R_{Q,2P})$, and 
    $\kappa_2(\UWTheta_Q)$, 
since $\kappa_1(\UWTheta_Q)$ and the covariance of $\R_{Q,2P}$ and $\UWTheta_Q$ are guaranteed to vanish by $\UWTheta_Q \rightarrow -\UWTheta_Q$ symmetry.
Higher-order contributions with $n\geq 3$ would vanish in the infinite statistics $N\rightarrow \infty$ limit if $\R$ and $\UWTheta_Q$ were exactly normally distributed and independent.
If an unwrapping algorithm can be chosen so that this is approximately true, fast convergence can be expected in the cumulant series.
In practice, the series must be truncated at some finite order $n_{max}$. Systematic uncertainties can be assigned by comparing results for $\widetilde{E}^{(n_{max})}$ with multiple truncation points $n_{max}$.

The correct ground-state energy $E_Q \equiv E_{Q,0}$ is reproduced at second-order in the expansion if the variance of $\widetilde{\Theta}_Q$ is approximately $1/\kappa_Q \approx 2 E_Q$.
The StN results of Eq.~\eqref{eq:StNMbartilde} can therefore be applied to $\widetilde{G}_{Q,2P}^{(n_{max})}$ if $E$ is replaced by $E_Q$ to give
\begin{equation}
  \begin{split}
    \text{StN}\left( \widetilde{G}_{Q,2P}^{(2)} \right) &\approx \frac{\sqrt{N}}{2E_Q t}\left[ 1 + O(N^{-1}) \right].
  \end{split}\label{eq:EQPStN}
\end{equation}
Despite the vanishing of all cumulants with $n\geq 3$ under the assumptions of
Eq.~\eqref{eq:small},
the statistical uncertainties of these higher cumulants increase with increasing $n$.
For large $n$, the variance of the $n$th cumulant will be dominated by the variance of the $n$th moment,
leading to StN behavior
\begin{equation}
  \begin{split}
    \text{StN}\paren{ \frac{\kappa_n}{n!} } \approx
    \text{StN}\left( \frac{1}{(2n!)}\widetilde{\Theta}^{2n} \right) \approx \sqrt{N} 2^{-n + 1/4} \left[ 1 + O(n^{-1}) + O(N^{-1}) \right].
  \end{split}\label{eq:momStN}
\end{equation}
At higher orders, the scaling of the StN with $E_Q t$ is not immediately analytically tractable, but it is not expected to fall exponentially. Eqs.~\eqref{eq:EQPStN} and \eqref{eq:momStN} then suggest that, in the small fluctuation approximation of Eq.~\eqref{eq:small}, StN ratios for $\widetilde{G}_{Q,2P}^{(n_{max})}$ decrease subexponentially with increasing $E_Q  t$ but exponentially with increasing $n_{max}$. These expectations are confirmed numerically in Sec.~\ref{sec:unwrap1D}.

Eq.~\eqref{eq:momStN} shows that even under the simplifying assumptions of Eq.~\eqref{eq:small}, the construction of a complete solution to the sign problem using phase unwrapping and the cumulant expansion still requires an extrapolation $n_{max}\rightarrow\infty$ where $N$ must be taken exponentially large in $n_{max}$ to remove all truncation errors at fixed statistical precision.
For LQFTs including LQCD, observations of the ubiquity of complex-log-normally distributed correlation functions~\cite{Hamber:1983vu,Guagnelli:1990jb,Endres:2011jm,DeGrand:2012ik,Drut:2015uua,Porter:2016vry,Wagman:2016bam} suggest that useful results might be obtained using modest $n_{max}$ despite the exponential difficulty of extrapolating to $n_{max}\rightarrow \infty$.
Understanding the size of truncation errors in practical calculations and systematic limitations of this method will likely require specific studies for particular LQFTs of interest.

\section{Monte Carlo results}\label{sec:unwrap1D}
MC simulations of $(0+1)D$ scalar field theory enable direct analysis of the efficacy of using cumulant expansions based on unwrapped moments to address the StN problem in charged correlators. The assumptions of Eq.~\eqref{eq:small} are dropped and fully general discussion of the free and interacting cases is presented below.
To produce unwrapped phases for the cumulant analysis, three numerical unwrapped schemes are considered that each satisfy $W[\widetilde{\theta}] = \theta$ but enforce different smoothness criteria to define $\widetilde{\theta}$,
\begin{enumerate}
\item Single-point integration: enforce 
$\abs{\widetilde{\theta}(t) - \widetilde{\theta}(t-1)} < \pi$ as in Eq.~\eqref{eq:1Dbound}.
\item Windowed integration (window $w$): enforce
$ \abs{\widetilde{\theta}(t) - \frac{1}{\text{min}(w,t)} \sum_{t' = \text{max}(t- w, 0)}^{t-1} \widetilde{\theta}(t')} < \pi.$
\item Gaussian-weighted integration (width $\sigma$): enforce 
$\abs{\widetilde{\theta}(t) - \sum_{t' = 0}^{t-1} \mathcal{N} e^{-(t'-t)^2/(2 \sigma^2)} \widetilde{\theta}(t')}$ $ < \pi$ with normalization $\mathcal{N}$ fixed by $\sum_{t' = 0}^{t-1} \mathcal{N} e^{-(t'-t)^2/(2 \sigma^2)} = 1$.

\end{enumerate}
The second and third smoothness criteria enable more robustly handling large phase
fluctuations by averaging a local neighborhood of phases and enforcing smoothness
on larger scales.

Families of MC ensembles $A,B,C,D,E$ were generated with five different values of $|M^2|$, corresponding to lattice spacings that range from very coarse to very fine while $L$ was scaled to hold physical lattice extent fixed. Families $A$ ($|M^2| = 0.1$), $B$ ($|M^2| = 0.025$), and $C$ ($|M^2| = 0.00625$) were used for a full spectral analysis, while $D$ ($|M^2| = 0.0015625$) and $E$ ($|M^2| = 0.000390625$) were included to determine the effect of lattice spacing on phase unwrapping. Within each family $X$, both a non-interacting ensemble (denoted $X_0$) and interacting ensembles (denoted $X_n$) were generated with $n \in \curly{1,2}$ indicating which of two dimensionless couplings $\lambda L / |M^2|$ were used. When $\lambda > 0$, a negative-mass phase can be accessed by choosing $M^2 < 0$. A further superscript on interacting ensembles ($X_n^\pm$) denotes the sign of $M^2$. 
MC ensembles were also generated using the dual form of the theory in which phase fluctuations are analytically integrated out in order to obtain precise results for verifying the accuracy of phase unwrapping; see Ref.~\cite{Detmold:PhaseUnwrapping} for details.

\begin{figure}[t]
\centering

\vspace{0.5cm}
\includegraphics[width=\textwidth]{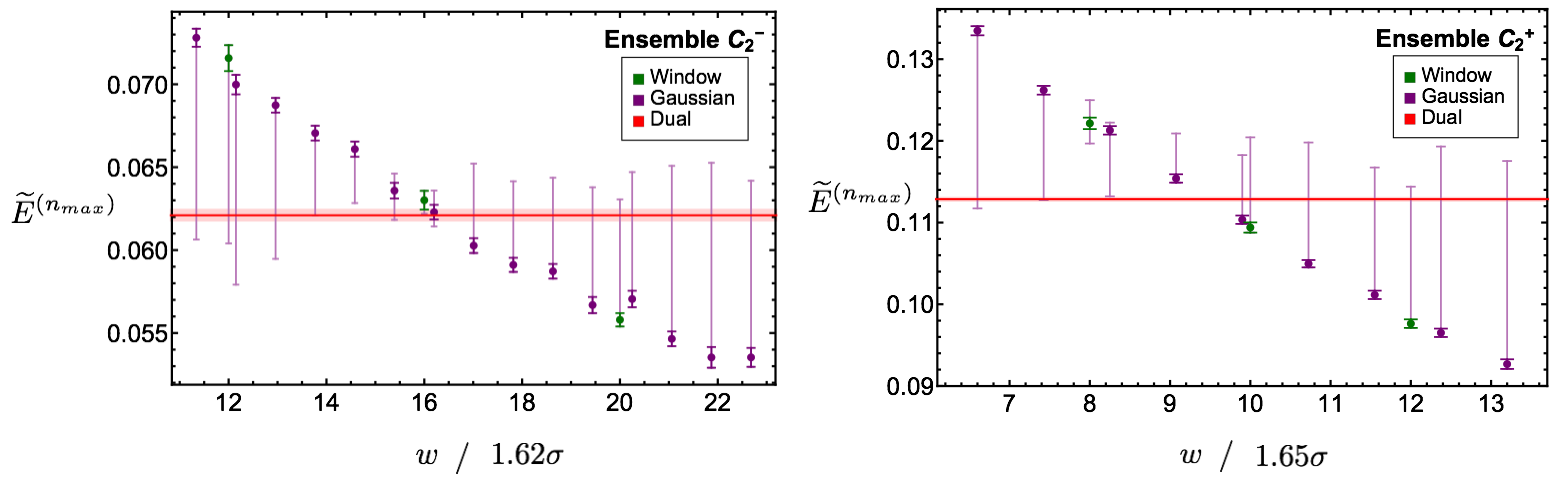}
\caption{Scalar boson mass estimates for ensembles $C_2^\pm$ versus window and width parameters in the windowed and Gaussian unwrapping schemes. An $O(1)$ constant of proportionality is estimated in each case to produce results consistent between the two schemes. Darker error bars indicate statistical uncertainty, while lighter errors bars include truncation errors estimated by the variation in central values of $\widetilde{E}^{(2)}$, $\widetilde{E}^{(4)}$, and $\widetilde{E}^{(6)}$. The red bands show dual ensemble results for comparison.}

\label{fig:gauss-window-sizes}

\end{figure}

\begin{figure}[t]
  \centering
\includegraphics[width=\textwidth]{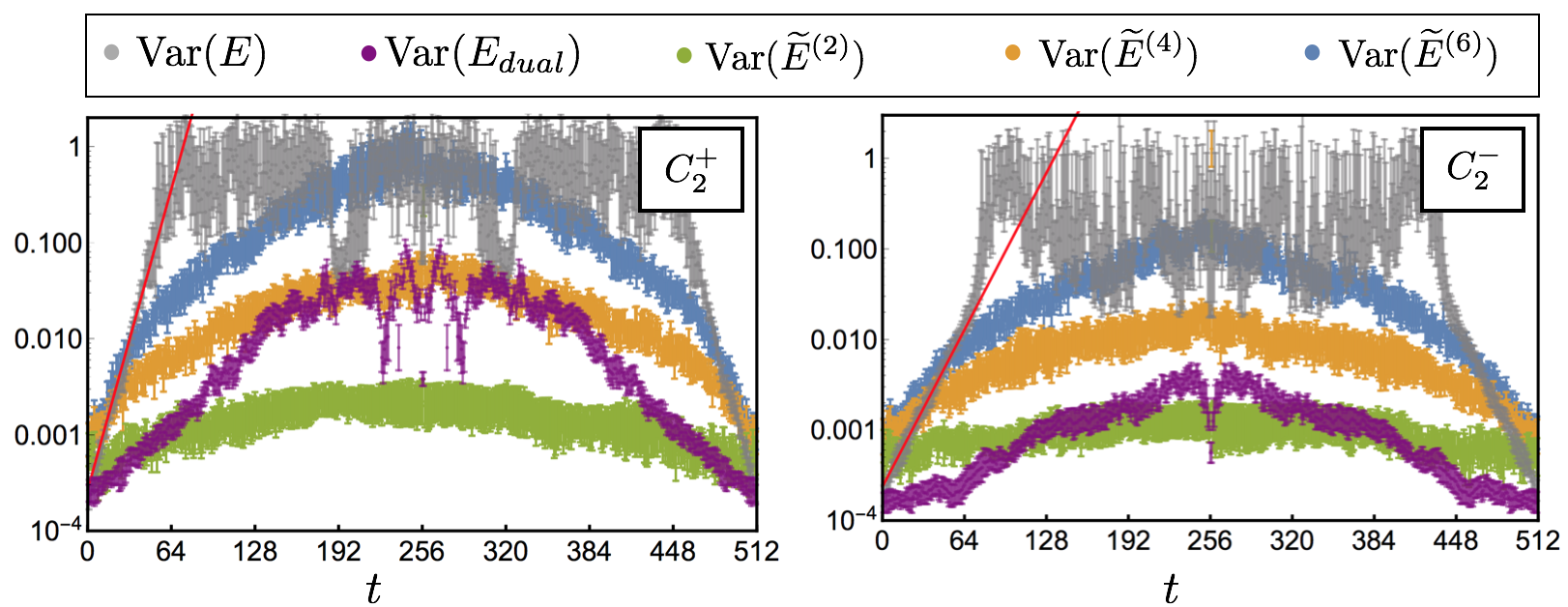}
\caption{Statistical variance in ground-state energy estimates versus correlator time separation for interacting ensembles $C_2^\pm$. The gray overlay plots the variance for the standard effective mass estimator, with
  a fit to the theoretical Parisi-Lepage estimate of exponential decrease $\mathcal{N} e^{-Et}$ in red.
  The purple points show the variance of the effective mass in the dual lattice variable ensemble and demonstrate exponential variance growth that is significantly less severe than the standard effective mass. The $n_{max} = 2$ estimate (green) with phase unwrapping has even less severe variance growth and becomes more precise than the dual variable estimate at large $t$. Phase unwrapped cumulant effective masses with $n_{max} = 2,\ 4,\ 6$ show variance growth with downward curvature on the logarithmic scale that is consistent with polynomial variance growth.}
\label{fig:cho-stn}
\end{figure}

A second-order cumulant estimate of the scalar boson mass using single-point integration is found to overestimate the true value on ensembles $A$, $B$, and $C$. 
Conversely, accurate results can be obtained from a second-order cumulant estimate using windowed or Gaussian-weighted integration with an appropriately chosen window or width. Fig.~\ref{fig:gauss-window-sizes} demonstrates such a tuning procedure for the windowed and Gaussian integration schemes applied to the scalar boson mass estimate $\widetilde{E}^{(n)}$ for representative interacting ensembles $C_2^\pm$. Systematic errors due to truncation are estimated from variation in the central values of higher-order truncations $n_{max} = 4, 6$ and are plotted in the figure using lighter error bars. When window size is smaller than the optimal size, the cumulant estimate converges from above, with higher order truncations becoming steadily more accurate at the cost of statistical precision. Conversely, larger window sizes show convergence from below with similar error scaling.
 Directional convergence can intuitively be explained by the effect of window size (or width) on the second-order variance estimate. Overly small window sizes are sensitive to lattice fluctuations below the physical scale and overly large window sizes prevent diffusion away from the mean, producing the over- and under-estimates of Fig.~\ref{fig:gauss-window-sizes}.
 Estimating truncation errors at multiple window sizes thus allows minimization of these errors by a good choice of phase unwrapping scheme, to the extent that error estimates are reliable.

The StN scaling in time separation is plotted in Fig.~\ref{fig:cho-stn} for the boson mass estimate using a tuned Gaussian integration scheme. The second-order cumulant contribution has almost constant StN at all time separations, and is more precise than the dual-variables comparison at large time separations. The StN for higher-order contributions decreases with time but at a rate slower than the standard mass estimator.

\begin{figure}[t]
\centering
\includegraphics[width=\textwidth]{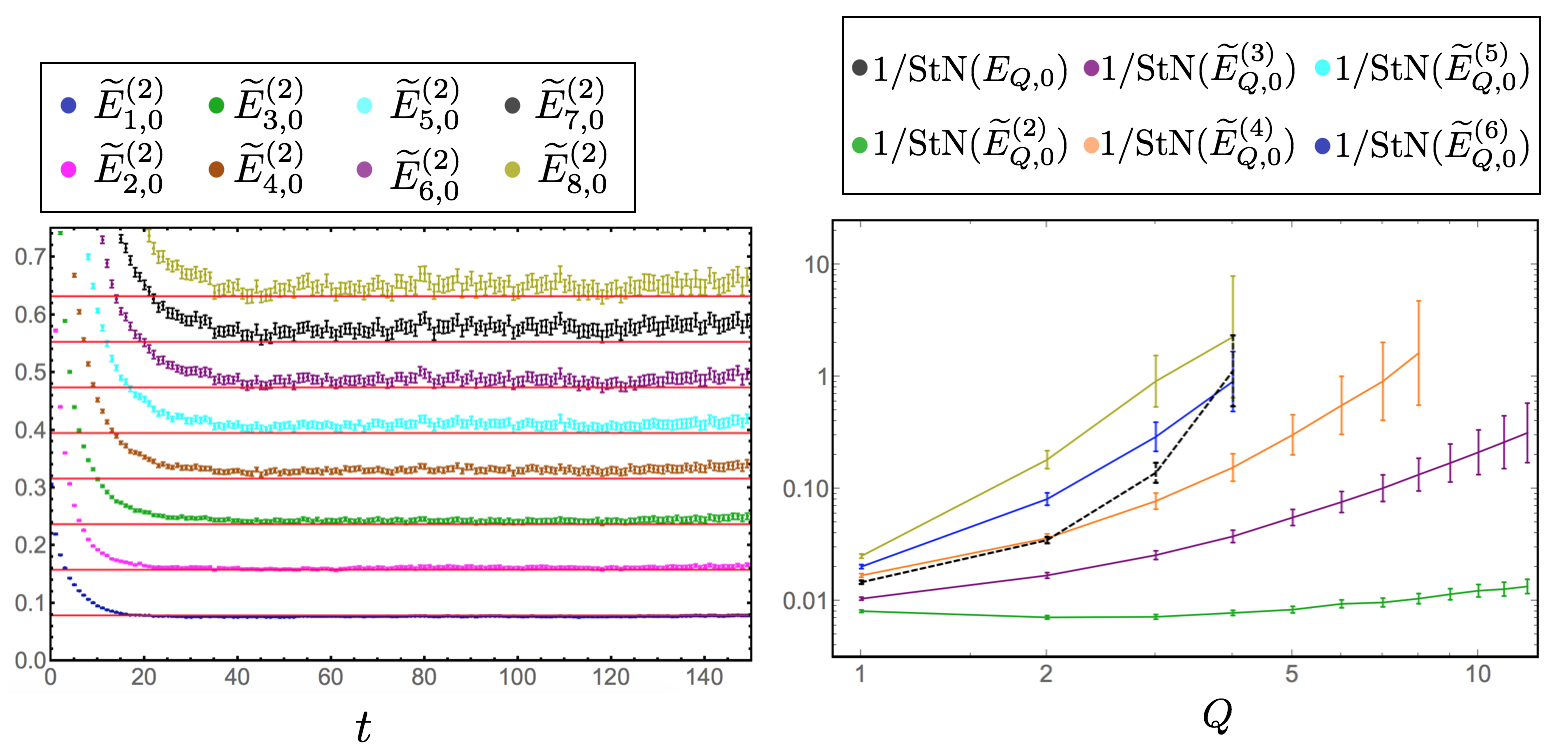}
\caption{ The left plot shows the second-order cumulant estimates of ground-state energies $\widetilde{E}_{Q,0}^{(2)}$ of charge sectors $Q=1,\dots,8$ for non-interacting ensemble $C_0$.
  The right plot shows the average inverse StN of these ground-state energy measurements for a time region $t = 10\rightarrow 20$ as a function of $Q$ for various cumulant expansion truncation orders. Gaussian-weighted integration with $\sigma = 1.41$ is used to calculate the unwrapped phase.}
\label{fig:unwrapStNQ}
\end{figure}

In higher charge sectors, a similar tuning procedure produces precise results at second order in the cumulant expansion for ground state energies $E_{Q,0}$. The optimal window in each case was found to approximately match the correlation length in the corresponding charge sector $w_Q \sim 1/E_{Q,0}$. Second-order estimates based on a well-tuned Gaussian integration scheme for a range of charge sectors are plotted in Fig.~\ref{fig:unwrapStNQ}, demonstrating minimal degradation of StN in time. Estimates of the StN scaling with cumulant order and charge sector are further plotted in the figure, indicating that StN for the second-order cumulant estimate does not degrade with charge $Q$, while the StN for higher-order truncations falls off in $Q$ at a rate slower than the exponential scaling of the standard estimator. The numerical results are thus compatible with the analytical results derived in Sec.~\ref{sec:scalarstats} which indicated subexponential scaling in $Q$ and $t$ for all truncations, but poor scaling with cumulant order $n$. Similar scaling is observed for interacting ensembles, with a full comparison of the low-lying spectra presented in Ref.~\cite{Detmold:PhaseUnwrapping}.
The results for both $Q=1$ mass estimates and higher-$Q$ correlators indicate that for second-order cumulant estimates, statistical precision is not a limiting factor. Instead, the systematic uncertainty associated with truncation dominates.

\begin{figure}[t]
  \centering
  \includegraphics[width=.9\textwidth]{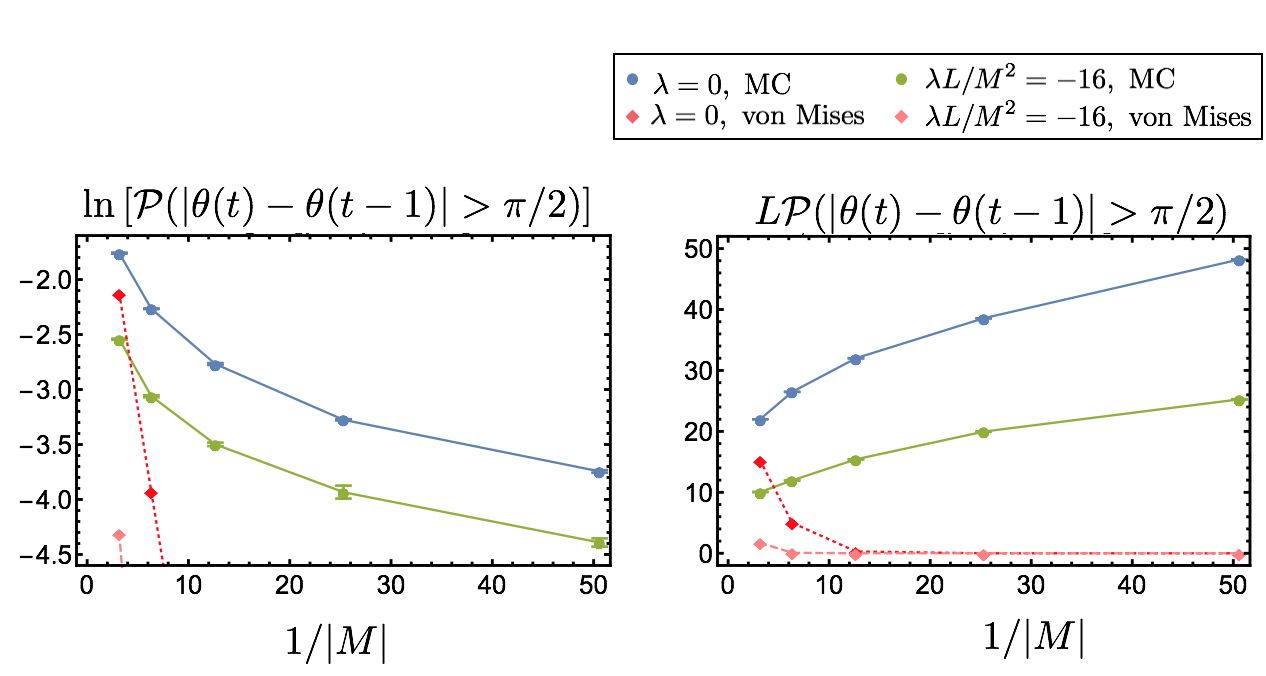}
  \caption{The left plot shows the probability of large phase jumps defined by $|\theta(t) - \theta(t-1)| > \pi/2$ for a variety of ensembles. The blue curve shows results for free-field ensembles $A_0$, $B_0$, $C_0$, $D_0$, and $E_0$ as a function of $1/|M|$ and therefore approximately as a function of the correlation length. The green curve shows analogous results for interacting scalar field ensembles $A_1^-$, $B_1^-$, $C_1^-$, $D_1^-$ and $E_1^-$ with $M^2 < 0$ and fixed $-\lambda L/M^2 = 16$. 
    The dotted red (dashed pink) curve shows the analytic small-fluctuation predictions corresponding to von Mises distributed phase differences with $\kappa \approx 1/(2E)$ calculated for the free (interacting) ensembles. The right plot shows the same probabilities multiplied by the lattice size $L$ to represent the expected number of large phase jumps per field configuration.
}
\label{fig:phase-jumps}
\end{figure}

The analytical description of phase unwrapping in Sec.~\ref{sec:unwrappedstats} suggested that unwrapping scheme sensitivity arose due to large phase differences between sites. For smaller lattice spacing, one might expect the frequency of such large jumps to decrease such that scheme dependence could be reduced as the continuum limit was approached. In the analytic approximation considered above, the probability of large jumps
$\mathcal{P}(|\partial_t \theta| > \pi - \varepsilon)$
can be calculated using the von Mises distribution.
For fixed $\varepsilon > 0$ the large jump probability then falls off approximately exponentially.

Correlations between $|\varphi|$ and $\theta$ break the assumptions leading to this analytic result. Fig.~\ref{fig:phase-jumps} compares the analytic expectation with numeric results. Numerical results show  that the probability of large phase differences actually falls off much slower than exponentially, and the probability that at least one link will break the small-difference assumption in fact grows. The scaling is surprisingly similar between free-field ensembles and interacting ensembles in the negative mass phase, where the average $|\varphi|$ is non-zero and near-zeros of the magnitude are expected to be infrequent.

This suggests a bleak future for phase unwrapping techniques in $(0+1)D$ scalar field theory, as the scheme sensitivity is not systematically improved in any limit of the theory. In general, $(0+1)D$ theories are particularly sensitive to phase unwrapping scheme due to the accumulation-of-errors described in Sec.~\ref{sec:unwrappedstats}: estimates for correlation functions are not only sensitive to scheme in local neighborhoods of large phase differences, but also at each subsequent point in the integration path. In a one-dimensional lattice, there is only one such integration path, forcing all integration to later times to go through regions of large phase fluctuation and leading to a variation of order $(2\pi)^n$ in the $n$th-order cumulant based on unwrapping scheme.

\section{Conclusions and future directions}\label{sec:conclusions}
Correlation functions possessing non-zero $U(1)$ charge face sign problems arising from phase fluctuations. Sign problems appear even in theories for which the vacuum partition function is positive-definite, including both free scalar field theory and QCD. This work demonstrates that an improved MC estimator for these correlation functions can be constructed, which trades the StN problems of the sample mean for truncation error in a cumulant expansion. Producing a convergent cumulant expansion relies on unwrapping the correlator phase $\arg(C(t))$ on each MC sample.
In a $(0+1)D$ toy model, analytical methods estimate that StN decreases polynomially in time separation at fixed cumulant order and numerical results confirm this expectation. 
While approximate analytical results suggest reduced unwrapping scheme dependence at finer lattice spacings, full numerical results indicate the opposite in one dimension.

The difficulties of phase unwrapping in $(0+1)D$ theories suggest directions for future work. In other fields, unwrapping in two~\cite{Goldstein:1988,Huntley:1989} and higher~\cite{Huntley:01,Hooper:07,Abdul-Rahman:09} dimensions has been shown to solve the accumulation-of-errors issue. It is the subject of future work to determine whether unwrapping of LQFT correlation function samples can be similarly improved by moving to higher dimensions.
More generally, cumulant expansion of log magnitudes and unwrapped phases provides a tunable technique to make a positive-definite estimate of correlation functions with non-zero $U(1)$ charge. Another promising direction for future work is to use such an approximation within ensemble generation to improve the StN characteristics of reweighting factors for charged correlation functions.

\vspace*{5mm}

{\bf Acknowledgments:}
	We thank Adam Bene Watts, Dorota Grabowska, David Kaplan, Christopher Monahan, Andrew Pochinsky, Martin Savage, Phiala Shanahan, Daniel Trewartha for helpful discussions.
         This work was partially supported by the U.~S.~Department of Energy through Early Career Research Award No.~de-sc0010495 and Grant No.~de-sc0011090 and by the SciDAC4 Grant No.~de-sc0018121.
        MLW was supported by a MIT Pappalardo Fellowship. All figures are reproduced from Ref.~\cite{Detmold:PhaseUnwrapping}.
	
\bibliographystyle{JHEP}
\bibliography{refs}

\end{document}

%% file: defines.tex
\usepackage{amssymb,latexsym}
\usepackage{slashed,array}
\usepackage{mathtools,amsfonts}
\usepackage{multirow}
\usepackage{bm,color,subfigure}

\newcommand{\avg}[1]{\left< #1 \right>} 

\newcommand{\R}{\mathcal{R}}
\newcommand{\UWTheta}{\widetilde{\Theta}}
\renewcommand{\TODO}[1]{{\color{red} #1}}